# Phase behaviour of deionized binary mixtures of charged colloidal spheres

**Nina J Lorenz[*], Hans Joachim Schöpe[*], Holger Reiber[*] and Thomas Palberg[*]**

[*] Institut für Physik, Johannes Gutenberg Universität Mainz, Staudinger Weg 7, D-55128 Mainz, Germany

**Patrick Wette[†], Ina Klassen[†], Dirk Holland-Moritz[†] and Dieter Herlach[†]**

[†] Institut für Materialphysik im Weltraum, Deutsches Zentrum für Luft- und Raumfahrt, Linder Höhe, 51170 Köln, Germany

**Tsuneo Okubo[‡]**

[‡] Institute for Colloidal Organization, Hatoyama 3-1-112 Uji, Kyoto 611-0012 Japan

We review recent work on the phase behaviour of binary charged sphere mixtures as a function of particle concentration and composition. Both size ratios $\Gamma$ and charge ratios $\Lambda$ are varied over a wide range. By contrast to hard spheres the long ranged Coulomb interaction stabilizes the crystal phase at low particle concentrations and shifts the occurrence of amorphous solids to particle concentrations considerably larger than the freezing concentration. Depending on $\Gamma$ and $\Lambda$ we observe upper azeotrope, spindle, lower azeotrope and eutectic types of phase diagrams, all known well from metal systems. Most solids are of body centred cubic structure. Occasionally stoichiometric compounds are formed at large particle concentrations. For very low $\Gamma$ entropic effects dominate and induce a fluid-fluid phase separation. As for charged spheres also the charge ratio $\Lambda$ is decisive for the type of phase behaviour, future experiments with charge variable silica spheres are suggested.

**Introduction**

The phase behaviour of one-component colloidal spheres has been studied extensively [1, 2, 3, 4, 5, 6]. Due to the tuneability of particle interactions in charged colloidal systems [7] a rich variety of crystal structures is observed ranging from noble gas like dense packing in hard spheres to plasma like open structures in systems with soft repulsion. At sufficiently large particle concentrations a kinetic glass transition is observed [8]. Also attractive terms may be introduced into the potential of mean force by addition of a second smaller component, typically a non adsorbing polymer [9, 10, 11]. Here the phase behaviour is further enriched and stable gas, liquid and crystal phases, but also metastable gels, clusters and an attractive glass are observed [12, 13, 14, 15].

Recently, interest shifted towards mixtures, where several additional parameters become important for the phase behaviour and system properties. Most important are the size ratio $\Gamma = a_S / a_L$ (where $a$ denotes the particle radius and the suffixes S and L refer to the small and large component, respectively), the charge ratio $\Lambda = Z_{eff,S} / Z_{eff,L}$ (where $Z_{eff}$ denotes the effective charge of each pure species) and the composition $p = n_S / (n_S + n_L)$ (where $n$ denotes the particle number density). Two types of mixtures have attracted particular interest. First, hard sphere or strongly screened charged sphere mixtures, where the phase behaviour is dominated by $\Gamma$. These form either stoichiometric compounds or amorphous states [16, 17, 18, 19, 20, 21, 22, 23, 24]. Second, oppositely charged spheres, where $\Lambda < 0$, Coulomb attraction dominates and the phase behaviour is again strongly influenced by packing constraints. Here a wealth of salt structures is found [25, 26]. A third class of systems has attracted much less attention so far: Data on mixtures of charged spheres under conditions where the repulsions are long ranged are still rare. The present contribution is devoted to these systems, where $\Lambda$ is positive. The charge ratio typically is close to the size ratio: $\Lambda \cong \Gamma$. This is due to counter-ion condensation occurring at sufficiently large bare charge, which causes the effective particle charge to be proportional to the particle radius [27, 28, 29]. The charge ratio may, however, in principle be varied over a large range by mixing high and low-charge spheres. This is here exploited to increase the miscibility in the crystalline phase and induce an upper azeotrope. Throughout the paper we will compare our findings to those made in hard spheres of similar size ratio and emphasize the influence of charges on the location of phase transitions and the general shape of the phase diagram.

For hard sphere mixtures early theoretical work suggested the formation of zero miscibility eutectic phase diagrams [30]. Later cell model calculations, computer simulations and experiments revealed a low, but finite miscibility leading to the formation of compounds at certain size ratios and compositions with peritectic and other types of more complicated phase behaviour [31, 32, 33, 34, 35, 36, 37]. Simple, eutectic phase behaviour is not any longer expected to dominate and in fact has not yet been seen in experiments on hard spheres. Substitutionally ordered solid solutions with spindle-type or azeotropic phase diagram are restricted to large size ratios $\Gamma > 0.8$ [34, 38]. The experimentally often observed vitrification is aided by the closeness of the kinetic glass transition, but recent calculations show that it is also connected to the unavoidable intrinsic polydispersity [39].

By contrast, charged sphere mixtures experience a soft repulsion, in particular under low salt conditions. Therefore, charged sphere mixtures are much less susceptible to size mismatch. This is true even for off-stoichiometric composition. Generally they are well miscible in both fluid and crystalline state. With increasing particle number density $n$ one observes stable fluids, substitutionally ordered solid solutions of bcc structure and finally of close packed crystal structure [40, 41, 42, 43, 44, 45]. In a few cases compound formation was observed at densities considerably larger than the melting density [43, 46, 47, 48]. Concerning the phase diagram type, Meller and Stavans [42] proposed that as a function of composition the shape of the phase boundaries in a $1/\Phi$ - $p$ representation varies with increased asymmetry from spindle-type over azeotropic to eutectic (where $\Phi$ denotes the volume fraction). This was connected to an increasingly lowered miscibility of the components in the solid state similar to the reasoning in hard spheres, atomic and molecular substances. They supported their conjecture by experimental phase diagrams showing a monotonically increasing freezing density at $\Gamma = 0.87$ and a peaked freezing at $\Gamma = 0.78$. At $\Gamma = 0.54$ two crystalline regions (of undetermined composition) were separated by an amorphous region, with the effect that the lower azeotropic or eutectic point was not reached. In a study on exhaustively deionized charged sphere mixtures Okubo and Fujita [43] reported a transition from spindle-type to azeotropic phase behaviour with coexisting fluids and solid solutions for size ratios decreasing from $\Gamma = 0.93$ to 0.77. In this study also glass formation was observed but only at elevated $n$. Systematic variation of composition for spindle-type phase behaviour is accompanied by a number weighted variation of other properties like conductivity, elasticity or solidification kinetics [44, 45], a concept originally introduced by Lindsay and Chaikin [40]. Finally, Kaplan et al. have shown that in charged but strongly screened mixtures of low to moderate asymmetry ($\Gamma \leq 0.4$) bulk and/or surface phase separation may occur, including the possibility of surface crystallization of the larger component [49].

The present paper goes beyond these studies, as it attempts to provide an empirical data collection based both on previous work and new results. It aims at providing an overview on the types of phase diagrams observable in deionized charged sphere suspensions. We will present the phase diagrams obtained in terms of particle number density and composition. Moreover we discuss the systematic influence of the different system parameters on the location of phase boundaries, compound formation and vitrification. Finally we will present a charge variable system. The one-component phase diagram of silica spheres depends on number density, effective charge and salt concentration. Silica spheres of varying charge may allow to adjust $\Lambda$ in a flexible way and further enhance the range of phase diagrams for charged sphere mixtures.

**Experimental**

We employed well characterized commercial colloidal spheres stabilized by sulphate, sulphonate or carboxylate surface groups (details of the synthesis are available from the manufacturers, see Table I) as well as home-made silica spheres synthesized in a Stöber process [50] and stabilized by silanol groups. Most particles have been used and characterized already in previous studies [51, 52, 53, 54, 55, 56, 57, 58, 59, 60]. The most important particle parameters are given in Table I. All particles were negatively charged. Charges quoted refer to the effective charge determined from elasticity measurements [55] testing the effective interaction potential and to the electrokinetic charge from conductivity measurements [61] testing the number of mobile counter-ions. The latter number is also obtained from cell model calculations [28, 62]. Previous studies have shown that in thoroughly deionized systems the electrokinetic charges are systematically larger than the effective charges by a constant factor of approximately 1.4 [44, 56]. This also holds for weakly charged particles [29]. The charge ratio therefore is not affected, when calculated from the same kind of charge. Sizes range from 68nm to 1103nm with standard deviations of 2-20%. PnBAPS68 and PnBAPS122 are kind gifts of BASF, Ludwigshafen, Germany. Si103 and Si136 were kindly donated by Catalyst & Chemicals Ind. Co., Tokyo, Japan. The main characteristics of the mixtures investigated in the present study, as well as those of [42] are compiled in Tab. II.

**Table 1:** Compilation of single particle properties: Lab code, with lab code used in previous studies in parenthesis and corresponding references indicated; diameter 2a, determined via Ultracentrifugation (UZ), Static light scattering in combination with Mie-theory (Mie), Dynamic light scattering (DLS) or Transmission Electron Microscopy (TEM), data given by the manufacturer are marked by an asterisk; polydispersity index (standard deviation normalized to the mean diameter); effective charge $Z_{eff}$ from elasticity measurements, electrokinetic charge $Z_\sigma$ from conductivity measurements or from cell model calculations (PBC) [28].

| Sample | Manufacturer & Lot# | 2a / nm | $\Delta a/a$ | $Z_{eff}$ | $Z_\sigma$ |
|---|---|---|---|---|---|
| PnBAPS68 [29, 51, 58] | BASF Batches ZK2168/7387 and GK0748/9378 | 68 (UZ) | 0.05 | 331±3 | 450±16 |
| Si84 [60] | Home made | 84 (Mie) | 0.08 | 253±15 to 340±20 | n/a |
| PS85 (D1B25) [43, 45] | IDC #767,1 | 85 (TEM)* | 0.07 | 350±20 | 530±32 |
| PS90 [45, 56] | Bangs Lab. #3012 | 90 (DLS)* | 0.025 | 315±8 | 504±35 |
| PS91 (D1C27) [51, 56] | Dow Chemicals #D1C27 | 91 (TEM)* | 0.07 | n/a | 680 (PBC) |
| PS100A | Bangs Lab. | 100 | 0.066 | 349±10 | 527±30 |
| [45] | #3512 | (DLS)* | | | |
| PS100B [45] | Bangs Lab. #3067 | 100 (DLS)* | 0.027 | 327±10 | 530±38 |
| Si103 (CS81) [51] | Catalyst & Chemicals Ind. Co. #CS81 | 103 (TEM)* | 0.13 | n/a | n/a |
| PS109A (D1P30) [51] | Seradyn Inc. #2010 M9R | 109 (TEM)* | 0.278 | n/a | 395±30 (PBC) |
| PS109B (D1B76) [51] | Seradyn Inc. #2011M9R | 109 (TEM) | 0.0028 | n/a | 450±30 (PBC) |
| PS120 [45, 54, 52] | IDC, #10-202-66.3 and #10-202-66.4 | 115 (Mie) | 0.016 | 474±10 | 685±10 |
| PnBAPS122 [58] | BASF #2035/7348 | 122 (UZ) | 0.02 | 582±18 | 743±40 |
| Si136 (CS121) [51] | Catalyst & Chemicals Ind. Co. #CS121 | 136 (TEM)* | 0.08 | n/a | n/a |
| PS1106 | IDC #PS2-1200, batch no. 2149 | 1106 (TEM)* | n/a | n/a | 8300±150 (PBC) |

**Table 2:** Compilation of properties of investigated mixtures: size ratio $\Gamma$; effective charge ratio $\Lambda$; type of the phase diagram observed: SP spindle, UA upper azeotrope, LA lower azeotrope, EU eutectic, PS fluid-fluid phase separation; $c$ residual small ion concentration; $n_G$ minimum density for the observation of an amorphous state; $n_M$ maximum melting density (minimum for PnBAPS68 - PS100B), $p(n_M)$ composition of the latter. The last column gives corresponding references. The last three mixtures are included from literature [42] for comparison.

| Samples | $\Gamma$ | $\Lambda$ | Type | $c$ / µmol l$^{-1}$ | $n_G$ / µm$^{-3}$ | $n_M$ / µm$^{-3}$ | $p(n_M)$ | Ref. |
|---|---|---|---|---|---|---|---|---|
| 1) PS85 - PS91 | 0.93 | 0.78 | SP or LA | <0.01 | - | 6.5 | 0.87±0.1 | [43] |
| 2) PS90 - PS100B | 0.9 | 0.96 | SP | <0.01 | - | 2.9 | 0.8±0.1 | [63] |
| 3) PS100A - PnBAPS122 | 0.82 | 0.56 | LA | <0.01 | - | 5.2 | 0.8±0.1 | |
| 4) PS85 - PS109A | 0.78 | 1.34 | LA or EU | <0.01 | 69 | 48 | 0.24 | [43] |
| 5) PS85 - PS109B | 0.78 | 1.17 | LA or EU | <0.01 | 86 | 40 | 0.53 | [43] |
| 6) Si103 - Si136 | 0.76 | n/a | SP | <0.01 | 34 | 0.81 | 1.0 | [43] |
| 7) PnBAPS68 - PS100B | 0.68 | 1.01 | UA | 0.2 | - | 0.37 | 0.2 | [64] |
| 8) PnBAPS68 - PnBAPS122 | 0.56 | 0.57 | EU | <0.01 | - | 26-43 | 0.82 | [59] |
| 9) PnBAPS68 - | 0.061 | n/a | PS | 11 | - | - | - |

| PS1106 | | | | | | | | |
|---|---|---|---|---|---|---|---|---|
| MS1 | 0.88 | 0.78 | SP | n/a | - | - | - | [42] |
| MS2 | 0.77 | 0.66 | LA | n/a | - | - | - | [42] |
| MS3 | 0.54 | 0.36 | LA or EU | n/a | - | - | - | [42] |

For all particle species pre-cleaned stock suspensions of precisely known particle number density were prepared following a protocol detailed elsewhere [45]. Samples of desired composition $p = n_S / (n_S + n_L)$ (molar or number fraction) were obtained by mixing small amounts of the stock suspensions with distilled water. They were deionized again using a batch procedure, following [43] with an improved protocol: For each mixture series of samples with different particle concentrations were prepared in 2ml cylindrical vials (Supelco, Bellefonte, PA, USA). The uncertainty $\Delta n/n$ from this dilution process is smaller than 10%. Where necessary, $n$ was determined with an uncertainty better than 5% by static light scattering. The relative uncertainty in composition is below 5%. A small amount of mixed bed ion exchange resin (Amberlite, Rohm& Haas, France) was introduced and the vials were sealed with an air tight Teflon® septum screw cap, carefully avoiding gas bubbles. Samples were placed cap down on a shelf to avoid uncontrolled salt gradients leading to accumulation at the IEX [65]. Under gentle daily stirring they were exhaustively deionized over several weeks. The theoretical limit of the residual ion concentration is given from the ion product of water: $[H^+][OH^-] = 10^{-14}$ mol$^2$ l$^{-2}$, where the proton concentration is set by the amount of counter-ions released by the particles: $[H^+] = n Z_\sigma / (1000 N_A)$, where $N_A$ is Avogadro's number and $n$ the density in units of m$^{-3}$ = $1*10^{-18}$μm$^{-3}$. For the typical cases investigated here this amounts to a residual small ion concentration $c_B = [OH^-] \leq 10^{-8}$ mol l$^{-1}$. We note that this concentration formally corresponds to a screening length of $\kappa^{-1} = 0.6$μm. The actual screening length is smaller, as also the released counter-ions have to be accounted for. Still, $\kappa d_{NN} > 1$ and at the nearest neighbour distance $d_{NN}$ the pair energy is much larger than the thermal energy $k_B T$ for all investigated systems. The residual small ion concentration cannot be directly measured in batch conditioning. A very sensitive marker of exhaustive deionization, however, is given by the size of the crystallites of a freshly shaken sample. As detailed elsewhere [66], the crystallite size strongly depends on the deionization state via the homogeneous nucleation rate density. The crystallite size decreases with decreasing residual ionic contaminations. A constant small crystallite size thus indicates completion of the deionization process. After two months the samples were homogenized for a last time and left to stand undisturbed on a vibration free shelf.

For the mixture 2) PS90/PS100B a continuous conditioning technique was employed [45], which is much faster, but has a residual ion concentration on the order of the self dissociation product of water. In particular this method allows obtaining an extremely narrow spaced series of different $n$ at constant $p$. It therefore allows to determine the freezing and melting points exactly [67]. For the mixture 9)

PnBAPS68/PS1106 we worked at $CO_2$ saturated conditions. The dissociation of dissolved $CO_2$ buffers a pH of 5.5 corresponding to a residual ion concentration of ca. 11µmol l$^{-1}$. In [42] the authors used a batch procedure, but their phase diagrams were measured already five days after mixing. We therefore estimate the residual ion concentration to be between that of the $CO_2$ saturated and the thoroughly deionized systems.

The mixture 9) PnBAPS68/PS1106 was also mixed from pre-cleaned stock suspensions, but at number fractions very close to unity. Under such conditions the effective pair potential is expected to contain attractive terms due to depletion interaction [9, 10]. The samples were investigated in flat capillary cells of 500µm height and 47x15mm$^2$ lateral extensions. Samples were inspected by eye but also by high resolution microscopy to monitor the distribution of large particles. In addition to homogeneous large particle distributions we also observed the formation of small, highly dynamic clusters of 3-20 particles. In the course of weeks most samples with clusters evolved a lateral phase separation into a large particle rich and a large particle depleted region. This was visible by the naked eye due to the different turbidity. The phase diagram shown below was taken after three months, when no further development was observed. A more detailed account of this system will be given elsewhere.

Phase diagrams of all other systems were obtained by visual inspection of the prepared and thoroughly deionized vials. The crystal structure was determined from standard static light scattering using a Debye Scherrer geometry. In most cases two or more peaks were visible to allow for Miller indexing and structure identification. Amorphous samples were identified by their finite shear rigidity and the simultaneous absence of crystal Bragg peaks. For crystalline systems the number density *n* is related to the scattering vector *q* at the Miller indexed peak positions as:

$$n_{bcc} = 2 \left( \frac{2\nu}{\lambda} \frac{1}{\sqrt{h^2+k^2+l^2}} \right)^3 \sin^3\left(\frac{\Theta}{2}\right) \qquad (1)$$

where $q = (4\pi\nu/\lambda)\sin(\Theta/2)$, with ν denoting the index of refraction of the suspension, λ the vacuum wave length of the laser light, and Θ the scattering angle, respectively. This expression was also used to evaluate the position of the principle peak of the fluid structure factor, as the fluid shows a bcc short range order close to the phase transition [68, 69].

**Results**

Phase diagrams of binary mixtures

In Figs. 1-8 we show the phase diagrams of our mixtures as plots of the inverse number density versus composition. This representation originally suggested by [42] facilitates a convenient comparison to the conventional temperature-composition or pressure-composition representation of atomic and molecular substances. In most cases the number density was varied over more than two orders of magnitude. All figures show the individual data points obtained, except for 7) (PnBAPS68/PS100B). Here

Phase behaviour of deionized binary charged sphere mixtures        8

we had a much higher data point density in 1/$n$ direction and directly show the melting and freezing points.

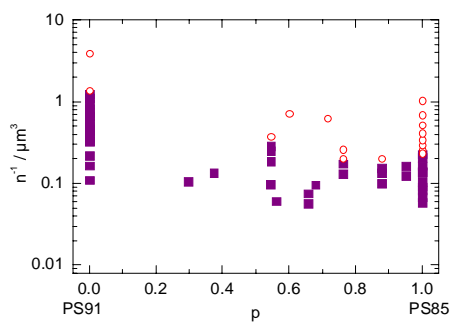

**Figure 1:** (colour online) Phase diagram of the mixture 1) PS85/PS91 obtained under exhaustively deionized conditions. Type of the phase diagram observed: Spindle or lower azeotrope. Symbols: open red circles: fluid phase, closed violet squares: crystal phase.

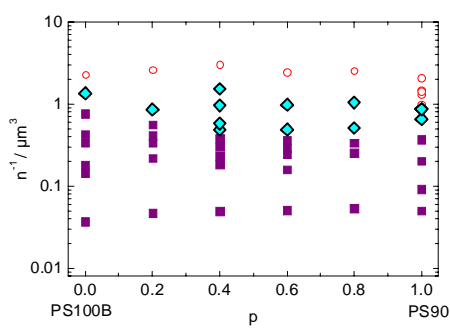

**Figure 2:** (colour online) Phase diagram of the mixture 2) PS90/PS100B obtained under exhaustively deionized conditions. Type of the phase diagram observed: Spindle. Symbols: open red circles: fluid phase, cyan shaded diamonds: coexistence, closed violet squares: crystal phase.

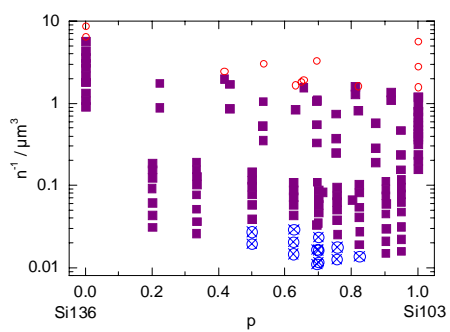

**Figure 3:** (colour online) Phase diagram of the mixture 6) Si136/Si103 obtained under exhaustively deionized conditions. Type of the phase diagram observed: Spindle. Symbols: open red circles: fluid phase, closed violet squares: crystal phase, crossed blue circles: amorphous phase.

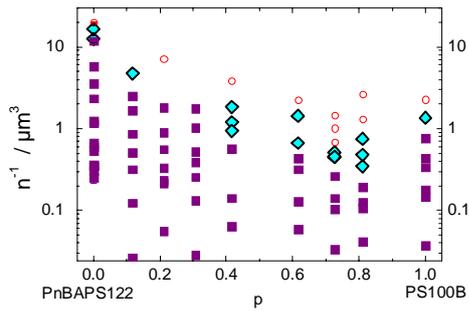

**Figure 4:** (colour online) Phase diagram of the mixture 3) PS100A/PnBAPS122 obtained under exhaustively deionized conditions. Type of the phase diagram observed: Lower azeotrope. Symbols: open red circles: fluid phase, cyan shaded diamonds: coexistence, closed violet squares: crystal phase.

The mixtures investigated show a rich variety of phase diagram types as a function of increasing size ratio. At low $\Gamma$ we observe spindle-type phase diagrams. The components of mixture 1), PS85 and PS91, with $\Gamma = 0.93$ differ in their freezing densities by almost an order of magnitude. The phase boundary in Figure 1 descends roughly linearly from the large value of PS91 and shows a shallow minimum around $p = 0.9$. The coexistence range was not explicitly determined for the mixtures using PS85. So we cannot decide whether this phase diagram is spindle or a shallow lower azeotrope. This is different for mixture 2) (PS90/PS100B) with $\Gamma = 0.9$ (Fig. 2). The pure species freezing points differ by about a factor of two and the coexistence region is almost horizontal. The upward bend of the freezing line (liquidus) and the downward bend of the melting line (solidus) are clearly seen. The coexistence region takes the characteristic spindle-type shape. A similar phase diagram is also seen in Figure 3 for mixture 6) (Si103/Si136) with $\Gamma = 0.76$. Here the pure species freezing points differ by roughly a factor of five. Interestingly, both pure species show high-lying freezing points in the $1/n – p$ representation, although both species have comparably large polydispersities of 0.08 and 0.13. The phase diagram of the mixture is of spindle-type. At the lower part of the phase diagram corresponding to large $n$ an amorphous region is found at intermediate compositions. On the low $p$ side the enhanced turbidity restricted the investigations. Investigations became unfeasible at somewhat lower densities and the existence of an amorphous phase could not be probed.

Mixture 3) (PS100B/PnBAPS122, Figure 4) with $\Gamma = 0.81$ again has strongly different freezing densities. The melting line descends roughly linearly and a more pronounced minimum is seen at about $p = 0.8$. A mixture at $p = 0.9$ is not yet accessible, so the precise location of the minimum is not yet determined. The phase diagram is classified as a lower azeotrope, as for all fully crystallized samples of PS100B/PnBAPS122 only a single crystalline phase was detected.

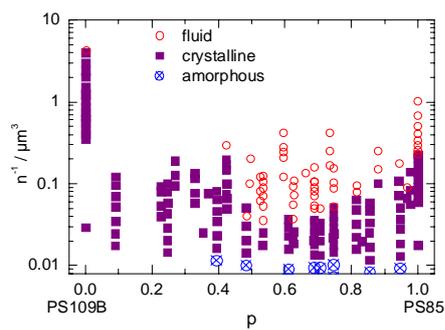

**Figure 5:** (colour online) Phase diagram of the mixture 5) PS85/PS109B obtained under exhaustively deionized conditions. Type of the phase diagram observed: Lower azeotrope or eutectic. Symbols: open red circles: fluid phase, closed violet squares: crystal phase, crossed blue circles: amorphous phase.

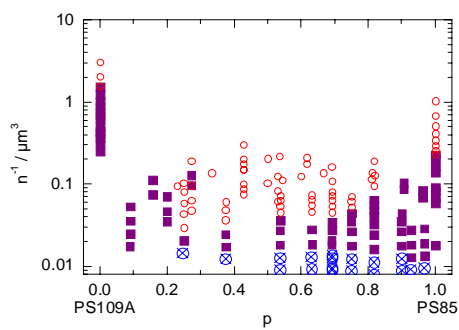

**Figure 6:** (colour online) Phase diagram of the mixture 4) PS85/PS109A obtained under exhaustively deionized conditions. Type of the phase diagram observed: Lower azeotrope or eutectic. Symbols: open red circles: fluid phase, closed violet squares: crystal phase, crossed blue circles: amorphous phase.

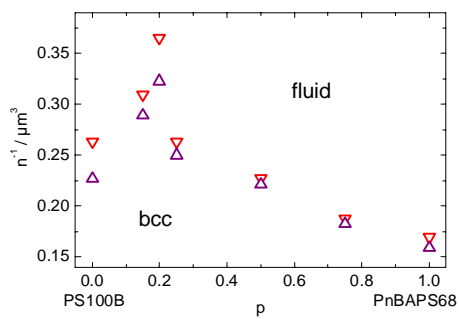

**Figure 7:** (colour online) Phase diagram of the mixture 7) PnBAPS68/PS100B obtained at low salt conditions. Type of the phase diagram observed: Upper azeotrope. Symbols: red down triangles: freez-

ing points, Violet up triangles: melting points. The region enclosed by the melting and freezing line denotes the coexistence of fluid and solid state.

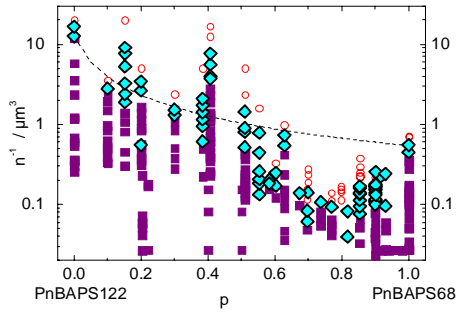

**Figure 8:** (colour online) Phase diagram of the mixture 8) PnBAPS68/PnBAPS122 obtained under exhaustively deionized conditions. Type of the phase diagram observed: Eutectic. Symbols: open red circles: fluid phase, cyan shaded diamonds: coexistence, closed violet squares: crystal phase.

In Figs. 5 and 6 we compare two mixtures at equal size ratio of 0.78. In both mixtures the smaller particles carry the larger charge. Mixture 5) (PS85/PS109B) has a charge ratio of $\Lambda = 0.99$ and PS109B a very low polydispersity and the lowest freezing density of all species. Mixture 4) (PS85/PS109A) has a charge ratio of $\Lambda = 1.13$ and PS109A a very large polydispersity. In both phase diagrams the crystal stability is strongly suppressed. In the first case the minimum is located on the small particle rich side. In the second case the region of suppressed crystal stability is extended also to large particle rich compositions. The minimum is more shallow and located at about $p = 0.2$. In both cases the difference in the number density at the minimum and at $p = 0$ amounts to two orders of magnitude. In both cases an amorphous phase is present at large $n$ over a larger range of compositions. For mixture 4) (PS85/PS109A) the density range over which crystalline samples were observed became very narrow in the region $p = 0.2$-$0.4$, while in mixture 5) (PS85/PS109B) crystals were detected over more than an order of magnitude in density. This effect was also present on the small particle rich side, but was less pronounced. For 5) (PS85/PS109B) the phase diagram is classified as either lower azeotrope or eutectic. The phase diagram of 4) (PS85/PS109A) in principle is similar but obviously strongly influenced by polydispersity. We therefore refrain from classification in terms coined for atomic substances.

In Figure 7 we show a particularly interesting sample. Mixture 7) (PnBAPS68/PS100B) with $\Gamma = 0.68$ and $\Lambda = 1$ showed an enhanced crystal stability at $p = 0.2$. The pure species freezing points differ by less than a factor of two, but between $p = 0.15$ and $p = 0.25$ the freezing points of the mixture lie even above that of PS100B. This sample was very carefully checked for its structure. At all compositions and densities which were covered in the present experiment we observed the formation of substitutional alloys of body centred cubic structure with no sign of compositional order. This phase diagram therefore is classified as an upper azeotrope.

The last phase diagram of crystallizing mixtures is shown for mixture 8) (PnBAPS68/PnBAPS122) in Figure 8. This mixture has $\Gamma = 0.56$ and $\Lambda = 0.57$, its pure sample freezing points are separated by one and a half orders of magnitude. We observe a complex shape of the phase boundary as a function of composition. At the small particle rich side a pronounced minimum is observed. At $p = 0.82$ the fully crystalline phase is not even reached within the range of accessible $n$. The dashed line gives the melting line estimate for a spindle-type phase diagram as derived from a linear interpolation between the $n_M$ of the pure samples. As compared to this line the suppression of crystal stability amounts to more than two orders of magnitude. As detailed elsewhere [70], the co-existence of two different crystal species could be demonstrated at $p = 0.9$ and $n = 43\mu m^{-3}$ by additional microscopic analysis in thin capillaries. Therefore the minimum at $p = 0.82$ is ascribed to a eutectic with the eutectic density located between $26\mu m^{-3}$ and $43\mu m^{-3}$. Interestingly, we observed congruent crystal formation also in the vicinity of the minimum, i.e. within experimental error the crystals were of the same composition as the melt. An additional comparably sharp feature of enhanced stability is visible at $p = 0.4$. Here the phase boundaries are above the dashed line.

Phase separation at very low $\Gamma$

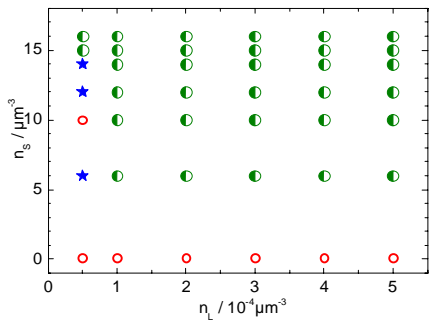

**Figure 9:** (colour online) Phase diagram of the mixture 9) PS1106/PnBAPS68 obtained at $CO_2$ saturation. Type of the phase diagram observed: fluid-fluid phase separation. Symbols: open red circles: fluid phase, stars: cluster phase, half filled circles: phase separation into a large particle rich and a large particle depleted fluid phase.

In Figure 9 we display the low $p$ part of the phase diagram of the mixture 9) PnBAPS68/PS1106 with $\Gamma = 0.06$. The number ratio in this mixture is about $p = 0.999$. Except for the lowest concentrations of large particles, this mixture shows phase separation into a large sphere rich phase and a large sphere depleted phase. Initially, small clusters of 3-20 particles were formed, which were highly mobile and showed vivid exchange dynamics with the surrounding monomers and other clusters. Phase separation occurred on a time scale of weeks via an increase of the cluster concentration. In Figure 9 the samples containing a spatially homogeneous distribution of clusters are marked by a star. One sample showed only occasional formation of transient clusters with sizes below 3 particles. In Figure 9 it is classified as fluid. Homogeneous fluid states were also observed for the pure PS1106 sample (bottom row). The dynamics and the detailed temporal development of this system is the subject of a forthcoming paper. The interesting point for the present paper is the occurrence of a phase separation at sufficiently large concentrations of small particles and the absence of kinetically arrested phases like gels or glasses.

Phase diagram types

The phase diagrams shown display a wide range of simple phase diagram types known from metals and other atomic substances. In Figure 10 we give an overview as a function of size ratio and charge ratio. Undecided cases are denoted by two symbols. For the mixture 6) (Si103/Si136) we assumed a proportionality of effective charge to particle radius [27]. Mixtures of particles of similar surface chemistry arrange close to the dashed line of slope one. With decreasing size ratio we observe a sequence of phase diagram types starting from spindle over lower azeotrope and eutectic to fluid-fluid phase separation. We also find an upper azeotrope with the miscibility in the crystalline phase increased above that in the fluid phase. From this survey we therefore may safely conclude that the miscibility of charged spheres decreases with decreasing size ratio. The influence of the charge ratio does not show a clear trend.

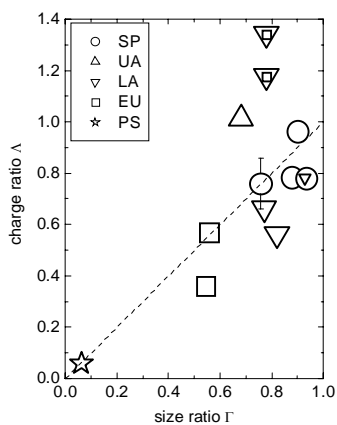

**Figure 10:** Compilation of phase diagram types observed. Circles: Spindle-type; down triangles: lower azeotrope; up triangles: upper azeotrope; squares: eutectic; Star: fluid-fluid phase separation. For the mixtures with uncertain assignment both symbols are given. The mixture with size ratio $\Gamma$ = 0.54 of [42] is regarded as eutectic, although the existence of two different crystal species here was impossible due to glass formation. Mixture 4) PS85/PS109A is denoted either lower azeotrope or eutectic, although it is actually not of a simple phase diagram type and the shape of the phase boundary is influenced by polydispersity. The mutual miscibility decreases with decreasing size ratio, while the role of the charge ratio remains to be clarified.

Polydispersity

Mixtures investigated were prepared from particles of significantly different polydispersity. PS109B has a very low spread of particle sizes. PS109A, PS85, Si103 and Si136 show the largest pure component polydispersities. Mixtures employing the latter particles show lower azeotrope or eutectic but also spindle-type phase diagrams. Moreover, the density needed for the formation of the amorphous phase is comparable. We therefore conclude that the influence of polydispersity on the phase diagram type is not very strong. However, mixing PS85 with the fairly monodisperse PS1009B yields a suppression of crystal stability mainly on the large $p$ side, while use of the strongly polydisperse PS109A suppresses the crystal stability over nearly the complete range of intermediate $p$. In this case the phase transition is rather close to the glass transition. Therefore a clear influence of polydispersity is seen concerning the location of the fluid-crystal phase transition. We note that also the addition of salt may destabilize the crystalline phase. In fact, the two lower size ratio mixtures of [42] both crystallized only at large particle concentrations for $p$ close to either zero or one and they showed glass formation directly from the melt at intermediate $p$.

Compound formation

Compounds were not in the focus of our investigation and no general structure determination was performed for densities further away from the phase boundary. However, we carefully checked the structure of our samples close to the melting line. In no case we found evidence for compound formation or compositional order. Rather all samples close to melting crystallize as substitutional alloys of body centred cubic structure. Compounds were only detected in a few cases at elevated $n$. Their structures observed are similar to those previously found in hard sphere or strongly screened systems [43, 47] and include $AB_2$ and $AB_{13}$ super lattices. To take up this point in more detail in future experiments, theoretical guidance would be very helpful to restrict experimental efforts.

**Discussion**

We have presented a collection of different phase diagram types formed in binary mixtures of charged colloidal spheres. Compared to hard sphere systems a main difference is the possibility to crystallize the systems already at fairly low densities and corresponding volume fractions $\Phi = n\,(4\pi/3)\,a^3$. This considerably extends the range of crystalline densities, before kinetic arrest in an amorphous state occurs. Also, compounds are not formed directly from the melt, which is the typical case in hard sphere systems. Rather, we observe the formation of substitutional alloys with no compositional order over one or two orders of magnitude in density. Only at densities significantly above the freezing density compound formation is observed and superlattice structures known from hard sphere systems are found here. At very low Γ we observe phase separation. This is known for neutral mixtures of hard spheres and non adsorbing polymers. However, in the absence of Coulomb repulsions, also a variety of kinetically arrested states like an attractive glass or gelation can be found [12]. The behaviour found here is also different to that visible in mixtures of charged particles with neutral depletant, where the formation of rigid finite size clusters is observed [71].

Another important result is the occurrence of regions with enhanced crystal stability at certain compositions. This was not yet observed in hard sphere systems, but is known from metal systems. There, enhanced crystal stability is observed either combined with the formation of a compound (which can be excluded here) or an upper azeotrope. The latter is less frequent than compounds but occasionally occurs as local maxima, e.g. in γ-AlMg or bcc AgMg or as global maxima, e.g. in AlMn [72]. In this respect and also concerning the suppressed tendency to form a glass already at the liquidus, the charged sphere mixtures resemble binary metal systems more strongly than the hard sphere mixtures studied previously. Also, the mixture 8) PnBAPS68/PnBAPS122 shows that the collection of phase diagrams is not restricted to very simple types, but that both regions of suppressed and enhanced crystal stability may occur simultaneously. The comparison in Figs. 5 and 6 on the other side shows that the intrinsic polydispersity is an issue for the actual location of phase boundaries. Polydispersity is absent in atomic materials. Future studies aiming at a comparison of phase behaviour between those systems should therefore employ samples with small polydispersity to acquire a precise phase boundary location.

We have seen a clear influence of the size ratio, but the influence of the charge polydispersity remains to be clarified. Intuitively we would expect a large miscibility in the crystalline phase at charge ratios close to unity. In fact, the mixture 7) of PnBAPS68/PS100B displays similar freezing densities for the pure components and a maximum in crystal stability at $p = 0.2$. Yet, at slightly larger charge ratios we again observe features expected for indifferent or reduced miscibility in the crystal as compared to the fluid. At this point we suggest investigating mixtures of one system with constant charge with another system of variable charge. To demonstrate the possibility of controlled charge variation we conducted experiments with Silica particles prepared by Stöber synthesis [73, 74]. The particles carry weakly acidic Silanol groups (Si-OH) on their surface, which partly start to dissociate in a deionized water environment. The surface charge density of the particles may be increased by adding NaOH [75] leading to the counter-ion exchange reaction: $SiOH + NaOH \rightarrow SiO^- Na^+ + H_2O$, until all surface groups are dissociated. Further addition of NaOH leads to a screening effect, which subsequently reduces the interaction between the particles.

In Figure 11 we display a recently determined phase diagram for a Silica system of $2a = 84$nm as a function of added salt concentration and particle number density. Here the continuous conditioning technique of [45] was used in combination with an automated titration. This yielded a large density of data points in a manageable time. A crystalline phase is observed for number densities above $n = $ $18\mu m^{-3}$. The crystalline region shows two fluid-solid phase boundaries. The boundary on the left is reached by charging up the particles at still deionized conditions. On the right it is reached by increased screening. The horizontal error bars denote the extension of the coexistence region. Recent measurements using small angle X-ray scattering indicate that this peculiar shape is also retained at larger $n$ [60].

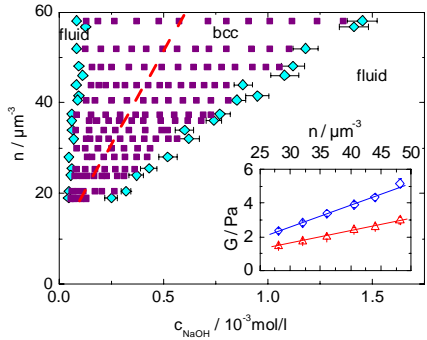

**Figure 11:** (colour online) Phase diagram of a Silanol-stabilized Silica species Si84 as a function of added NaOH and particle number density. Symbols: filled squares: crystalline phase; cyan diamonds: fluid-crystal coexistence with the extension of the coexistence region indicated by the horizontal error bars. The left phase boundary is reached as the spheres are charged up under deionized conditions. The dashed red line indicates conditions of maximum interaction reached at the equivalence point of the titration. Upon further addition of NaOH the interaction becomes strongly screened and the system melts again. The insert shows the shear moduli measured at different densities at the left phase boundary (up triangles) and under conditions of maximum interaction (diamonds). The lines are fits to the data yielding $Z_{eff} = 253\pm15$ at the left phase boundary and $Z_{eff} = 340\pm20$ at maximum interaction.

The red dashed line indicates the location of the maximum interaction as observed in measurements of the elasticity and of conductivity. The insert shows the shear moduli determined at the left phase boundary (down triangles) and at maximum interaction (diamonds). The lines are fits to the data for a body centred cubic structure [56, 60]. Irrespective of particle concentration we find an effective charge of 253±20 at the left phase boundary and of 340±20 at maximum interaction. This nicely demonstrates the possibility to tailor the interaction in a colloidal charged sphere system while staying under deionized condition. With a mixture of spheres stabilized by strongly acidic groups and Silica spheres of variable charge (staying at the left side of the line of maximum interaction) would allow experiments with constant Γ but varied Λ.

**Conclusions**

We have reviewed recent work on the phase diagram of binary mixtures of charged colloidal spheres, which complements previous work on hard sphere model systems. Under exhaustively deionized con-ditions the long ranged Coulomb repulsion considerably enhances the stability of the crystal phase and in general brings the phase diagram types observed closer to the cases known from metals. With decreasing size ratio we observed a rich variety of simple phase diagram types including spindle, upper and lower azeotrope, and eutectic but also more complex shapes. To further elucidate the still unexplained role of the charge ratio we suggested experiments using a charge variable species. The present investigation may well serve as first orientation on the kind of phase diagram to expect for a given size and charge ratio under deionized conditions. We further hope to have stimulated increased theoretical interest in this subject. Clearly, the computational challenge for charged systems by far exceeds that of systems with short ranged potentials. However, a deepened and theoretically well founded understanding of the similarities (and the differences) of the here presented binary phase diagrams to those of metals would be more than rewarding.

**Acknowledgements**

It is our pleasure to thank H. Fujita, H. Ishiki, J. Liu and A. Stipp for assistance with the experiments. We also thank K. Binder, H. Löwen, D. Holland-Moritz and J. Horbach for many fruitful discussions on alloy formation. This work was financially supported by the Japanese Research Foundation for Opto-Science and Technology, the Japanese Ministry of Education, Science and Culture, the Deutsche Forschungsgemeinschaft, within SPP1120 and 1296 (Pa459/12,14,16 and He HE1601/23,24), and the Materialwissenschafliches Forschungszentrum (MWFZ), Mainz. This is gratefully acknowledged.